\documentclass[]{aastex631}

\newcommand\moid{MOID}
\newcommand\sigmamoid{\sigma_{\moid}}
\newcommand\tof{ToF}
\newcommand\dtof{\Delta \tof}
\def\las{\mathrel{\hbox{\rlap{\hbox{\lower3pt\hbox{$\sim$}}}\hbox{\raise2pt\hbox{$<$}}}}}
\def\gas{\mathrel{\hbox{\rlap{\hbox{\lower3pt\hbox{$\sim$}}}\hbox{\raise2pt\hbox{$>$}}}}}

\begin{document}

\title{Encounter circumstances of asteroid 99942 Apophis with the catalogue of known asteroids}

\author{Paul Wiegert}
\affiliation{Department of Physics and Astronomy \\
The University of Western Ontario\\
London Ontario Canada}
\affiliation{Institute for Earth and Space Exploration \\
The University of Western Ontario\\
London Ontario Canada}

\author{Ben Hyatt}
\affiliation{Department of Physics and Astronomy \\
University of Waterloo\\
Waterloo Ontario Canada}



\begin{abstract}

Asteroid 99942 Apophis will pass near the Earth in April
2029. Expected to miss our planet by a safe margin, that could change
if Apophis’ path was perturbed by a collision with another asteroid in
the interim. Though the statistical chance of such a collision is
minuscule, the high risk associated with Apophis motivates us to
examine even this very unlikely scenario. In this work, we identify
encounters between known asteroids and Apophis up to April 2029. Here
we show that Apophis will encounter the 1300~m diameter asteroid 4544
Xanthus in December 2026. Their Minimum Orbit Intersection Distance
(MOID) is less than 10,000 km, with Xanthus passing that closest point
just four hours after Apophis. Though a direct collision is ruled out,
the encounter is close enough that material accompanying Xanthus (if
any) could strike Apophis. We also identify other asteroid encounters
that deserve monitoring.

\end{abstract}



\section{Introduction} \label{sec:intro}

Asteroid 99942 Apophis is one of the most studied near-Earth
asteroids, stemming in large part from its close passages to the
Earth. Initially thought to pose an impact danger, careful
observations of its trajectory and properties have shown the risk of a
collision with our planet during any of its upcoming close approaches
in 2029  and 2036 is
nil\footnote{\url{https://www.jpl.nasa.gov/news/nasa-analysis-earth-is-safe-from-asteroid-apophis-for-100-plus-years}}. However,
these assessments e.g. \cite{farchecho13,vokfarcap15} among many others, have assumed
that the motion of the asteroid will continue uninterrupted by any
impulsive perturbations such as might arise from Apophis' collision
with another asteroid.

The {\it a priori} probability of an asteroid colliding with Apophis
during the period of a few years between now and that asteroid's close
approach to Earth in 2029 is extremely low. Nevertheless, asteroid
collisions are believed to occur. Asteroid families are thought to
originate from a disruptive collision between asteroids
\citep{micbenpao01}. Also, the probable collision of smaller asteroids
with larger ones has been observed at least twice in the recent past:
P/2010 A2 is thought to have been struck in 2009 by an object a few
meters in size \citep{snotubvin10, jewweaaga10}, and asteroid 596
Scheila in 2011 by a 35 meter-sized body
\citep{ishhanhas11,jewweamut11}.

Even if a particular asteroid might not directly collide with Apophis,
material accompanying that asteroid (whether gravitationally bound or
not) could also pose some risk. Binary and multiple asteroids are seen
among all asteroid populations, including the near-Earth asteroids
\citep{marpratay15}. Solar system bodies may also shed material from
their surface through processes driven by water ice sublimation. A
comet is the classic example, and usually has a meteoroid stream
composed of material shed from its surface travelling along its
orbit. So-called 'active asteroids' may also release material through
a wide array of mechanisms: impact disruption, rotational
destabilization \citep{hsi17}, thermal fracturing \citep{knifitdel16},
radiation pressure sweeping and electrostatic levitation
\citep{jewhsiaga15}. Active asteroids may release substantial amounts
of mass.  For example, active asteroid 3200 Phaethon shows no or only
very weak cometary activity \citep{tabwieye19} and yet is accompanied
by a substantial debris stream (total mass $10^{13-14}$~kg
\citep{hugmcb89,bla17}) that produces one of the strongest meteor
showers at Earth, the Geminids meteor shower.

For completeness, we note that even in the absence of a collision,
there is a change in Apophis' velocity $\Delta v$ due to the
gravitational attraction of a passing asteroid. The magnitude of this
change can be estimated through the impulse approximation
\citep{aguwhi85}
  \begin{equation}
    \Delta v \approx \frac{2Gm}{v_{rel}d}  \label{eq:impulse_approx}
  \end{equation}
where $m$ is the mass of the passing body, $G$ is the gravitational
constant, $d$ is the minimum distance between the body and Apophis,
and $v_{rel}$ is the relative velocity at the time of their closest
approach. Such changes will invariably turn out to be
negligibly small for the cases examined here.

Though the chances of Apophis undergoing an interaction with another
asteroid that could affect its impact risk with the Earth is minimal,
the high stakes involved motivate us to extend the study of this
asteroid even to these unlikely eventualities. In this work we
identify close encounters between Apophis and members of the catalogue
of known asteroids and comets, with the purpose of
\begin{enumerate}
\item  examining the probability of a direct collision
\item assessing the possibility of collisions with asteroidal
  satellites or other accompanying debris
\item planning for telescopic observation of Apophis-asteroid
  encounter events to verify the absence of a collision and maintain
  situational awareness
\item identifying those asteroids for which additional observations or
  analysis are needed to better assess their collision probability
  with Apophis.
\end{enumerate}

\section{Methods} \label{sec:methods}

The primary goal of this analysis is to assess the risk of a known
small body in the Solar System impacting Apophis prior to its close
approach to Earth in April 2029. Here we examine the time frame
beginning 25 Feb 2023 (JD 2460000.5) and ending 13 Apr 2029 (JD
2462239.5) which is the time of Apophis' next close approach to
Earth. Though an examination of possible impacts with Apophis after
the 2029 close approach is also of interest, it will be more
appropriate once the deflection of Apophis from its current orbit by
that encounter has been measured precisely.

Two catalogues of asteroid and comet orbital information are used and
the results compared.  The Small-Body Database, maintained by NASA's
Jet Propulsion Laboratory (JPL) on their Solar System Dynamics (SSD)
web page\footnote{\url{https://ssd.jpl.nasa.gov/tools/sbdb_query.html}
retrieved 1 May 2023} contains the orbital elements for over 1.2
million asteroids and comets. The data retrieved from the JPL
Small-Body Database for the present study is current as of 1 May 2023.
The NEODyS-2 database sponsored by ESA contained information on 31886
near-Earth asteroids when retrieved on 1 May
2023\footnote{\url{https://newton.spacedys.com/neodys/index.php?pc=5},
retrieved on 1 May 2023}.

The orbital determinations of both the JPL and NeoDys teams are based
on the same database of observations maintained by the IAU's Minor
Planet Enter and employ similar techniques, yet their results differ
slightly. Each of these two teams of global experts accommodate the
inherent measurement uncertainties in asteroid and comet observations
via different approaches, informed by their collective expertise. This
independence serves as an effective mechanism for cross-validation,
and enhances the reliability of near-Earth asteroid catalogues. This
will also turn out to have some importance in the interpretation of
our results in rare cases where they disagree substantially, which we
will return to in Section~\ref{sec:conflict}.

\subsection{Initial sub-sample selection}

To properly assess the impact probability of a particular asteroid
with Apophis requires a full-scale simulation of its orbit through
time, including planetary and other perturbations. Though we will do
such an analysis for the most interesting objects, we first filter the
JPL catalogue to remove the many objects which cannot approach
Apophis. At this filtering stage, we will assume that the orbital
elements provided by the catalogue are approximately constant over
the time interval considered, though we will see that this is not
always the case, and assess this further in
Section~\ref{sec:simulatedsample} below.

We ignore objects in the JPL catalogue whose probability of impacting
Apophis cannot be effectively assessed, because of missing orbital
elements and/or no uncertainties provided. We do not at this stage
exclude objects because they have large orbital uncertainties. We will
address their uncertainties carefully at later stages when we will use
their orbital covariance matrices to determine the effect of
uncertainty on our understanding of their trajectories.  We do at this
stage however eliminate those objects which cannot reach Apophis,
because their perihelia and/or aphelia do not overlap. An object is
eliminated if its perihelion $q$ is not at least as small as Apophis's
aphelion $Q_{A}$, or its aphelion $Q$ is not at least as large as
Apophis's perihelion $q_{A}$ within $5\sigma$ or specifically, if
either of the conditions below are met:

\begin{eqnarray}
  q -5\sigma_q &>& Q_{A} + 5\sigma_{Q_A} \\
  Q + 5\sigma_Q&<& q_{A} - 5\sigma_{q_A} 
\end{eqnarray}
where $\sigma_q$ and $\sigma_Q$ represent the uncertainties in the
perihelia and aphelia of the asteroids in question. This reduces the
JPL sample to about 30,000 objects.  These together with the entire
NeoDys catalogue, are passed to the next step.

In our second filtering step, we examine the Minimum Orbit
Intersection Distance (MOID) of these objects with respect to Apophis.
The MOIDs are computed using a simple 2D minimization of the distance
between two orbits as a function of their true anomalies using
Powell's method \citep{preflateu86}, with a tolerance of
$10^{-6}$. This is a general purpose minimization routine, widely used
and considered robust against pathological cases. Though much slower
than purpose-built MOID routines such as that of \cite{balmik19},
we've compared our routine extensively against the BM2019
implementation and ours gives answers identical to within the
tolerance.

At this stage, we continue to take the orbital elements of the
asteroids to be constant. Apophis' orbit is so well known that we will
model it as a single particle on its nominal orbit. For all other
asteroids, a set of 100 clones are generated from the covariance
matrices. The MOID is computed between each of the 100 orbit clones
and the nominal orbit of Apophis. The standard deviation $\sigmamoid$
of this set of 100 MOID values is taken as representative of the
dispersion of the object's nominal MOID with respect to Apophis. The
accurate calculation of the uncertainty in the MOID is certainly more
involved than simply taking the standard deviation e.g. advanced
methods for doing so are presented by \cite{grotom07}. However we did
not adopt a more detailed calculation here because we do not require a
precise uncertainty but a simple dispersion, as it used here only to
determine the outer envelope for our sample.

We will include objects in our next subsample if they meet one of the
two criteria below.  These criteria are chosen iteratively, by
examining the results of our simulations (discussed further below) and
in particular by determining by how much the MOIDs of the near-Earth
asteroids change over the course of a few years. We found that the
typical variation in the MOIDs of the asteroids in this study was
$\las 0.001$~AU. These changes result from planetary perturbations,
with a handful of cases (involving encounters with Jupiter) reaching
changes in the MOID of nearly 0.01~AU. Based on these values, we
choose our selection criteria to be as follows.

\begin{enumerate}

\item Any object with a nominal MOID less than 0.001~AU is included
  regardless of its orbital uncertainty. 

\item Any objects which have a $\moid < 5\sigmamoid$ and $\moid <
  0.01$~AU will also be included in our sample. This condition ensures
  that the MOID is small and consistent within 5 standard deviations
  with zero. We note that, for a true Gaussian distribution, only one
  part in 3.5 million of the integrated probability is beyond 5
  standard deviations. The distribution of the MOIDs is not strictly
  Gaussian but our choice of 5 standard deviations mean that we
  exclude objects whose orbits are so uncertain that they have less
  than of order $10^{-6}$ chance of intersecting the orbit of Apophis.




\end{enumerate}

Because of the possibility of the MOIDs changing due to planetary
perturbations, our sample may not include all asteroids that could
approach Apophis. A full numerical simulation of all near-Earth
asteroids (NEAs), or at least the approximately 30,000 objects
identified by our earlier filtering, would be necessary to categorize
such a sample exhaustively. This study discusses the most
straightforward cases, but a full examination of possible encounters
will be presented in a future work.

Our final filtering reduces the sample to 376 objects from the JPL
catalogue, and 396 objects from the NeoDys catalogue, many of the same
objects appearing on both lists. Most are asteroids, seven are known
comets. There are 322 objects that appear on both lists, while 54
objects appear only on the JPL list, and 74 objects only on the NeoDys
list. The non-overlapping objects all lie near the threshold
boundaries, appearing just inside in one sample and just outside in
the other.  Such differences are to be expected, and precisely why we
considered both catalogues independently. Each of the objects on our
final JPL and NeoDys lists will undergo more rigorous further analysis
discussed in the next section.

\subsection{Full simulation of objects in subsample} \label{sec:simulatedsample}

For the objects in our filtered subsample, a full-scale numerical
integration of their motion within the Solar System is
performed. Apophis is represented by a single particle on its nominal
JPL orbit. Each of the other asteroids is represented by 2000 clones,
1000 clones each generated from the JPL CNEOS covariance matrices and
NeoDys covariance matrices for that object.

The clones are integrated with the RADAU \citep{eve79} algorithm
within a Solar System which includes the eight planets, with the Earth
and Moon combined into a single body at their barycenter. Initial
conditions are from the JPL DE405 ephemeris \citep{sta98}. The
integration runs from JDE 2460000.5 to JDE 2462239.5 with an external
time step of 10 days. The candidate asteroid clones and Apophis are
treated as massless test particles, and non-gravitational forces such
as Yarkovsky are ignored. We recognize that our model does not include
the full suite of perturbations as are, for example, included in the
JPL Horizons model. We will nonetheless find that our results are
highly compatible with theirs where we can compare them, which we will
do later in this section.

To determine the possibility of a collision, we examine the 1) the
Minimum Orbit Intersection Distance (MOID) between Apophis' orbit and
that of a known asteroid, and 2) the difference in time of flight to
the MOID $\dtof$. Here the MOID will always refer to the mutual MOID between
Apophis and some other asteroid, unless otherwise specified. The time
of flight ($\tof$) to the MOID is the time needed for an asteroid to
travel from its current position to the MOID. The difference in the
time of flight to the MOID of two asteroids $\dtof$ represents the
interval of time by which they miss both being at their mutual MOID at
the same time.  For example, if Apophis will reach its MOID with
asteroid X in $\tof=$~4 days, and X will reach its mutual MOID with
Apophis in $\tof_{X}=$ 7 days, then $\dtof_{X} = -3$~days. We adopt
the convention that $\dtof$ is negative if Apophis reaches the MOID
first. The condition of $\moid = \dtof =0$ corresponds to a collision
with Apophis.

We select our sample based on the $\moid - \dtof$ values for two reasons.
\begin{enumerate}

\item Material released at low velocity from a small Solar System body
  tends to disperse along its orbit, even if released with an
  isotropic velocity distribution relative to its parent. This is the
  well-known mechanism by which a comet creates a meteoroid stream
  which may stretch for hundreds of AU along its entire orbit. As a
  result, if any material is released by an object examined in this
  study, it is reasonable to assume that that material will have an
  overall orbit shaped much like its parent's and have a similar MOID
  but with a larger spread in $\dtof$. This simplifies our
  interpretation of scenarios involving possible collisions with asteroidal
  debris. Consider two asteroids, each of which pass the same minimum
  physical distance $R$ from Apophis, but where one has a large MOID
  and the other a small one. Any material released from the first one
  cannot approach Apophis to small distances, while material (if any)
  released from the asteroid with a small MOID may. Thus asteroids
  with small MOIDs naturally entail a higher risk of collision with
  any material that may have been shed by it.

\item A useful property of the MOID and $\dtof$ is that both are
  approximately constant during the days and weeks around a particular
  close approach between two asteroids. Using these quantities avoids
  difficulties associated with selecting close approaches between
  Apophis and another asteroid based on their mutual distance (which
  is rapidly changing) within a numerical simulation. Such a
  simulation, which inevitably produces output at some discrete set of
  time steps, could inadvertently step over a close encounter without
  careful interpolation, but our choice of quantities avoids this
  difficulty entirely.

\end{enumerate}

Our approach is similar to the use of the Opik target plane, e.g. as
described in \cite{fareggcho19}. The Opik target plane offers the
helpful properties that the minimum absolute value of the $x$-axis is
the MOID, and the $y$-axis is related to the timing offset, with $y=0$
indicating that both bodies arrive at their mutual MOID at the same
time. However, the Opik plane assumes that the closest point of
approach is near enough to the MOID that the motions of the bodies can
be taken to be linear during the encounter. This is not the case for
many of the encounters we examine, where an object with a small MOID
with Apophis may have a $\dtof$ of months or years. In our
formulation, both the MOID and the difference in time of flight to the
MOID always remain well-defined and intuitive, while the Opik target
plane does not.

We verified the details of the close approaches discussed below with
JPL's Horizon's integrator. We queried the SSD/CNEOS Horizons API
service for the nominal Cartesian positions and velocities of Apophis
and the asteroid in question during the time of the close encounter,
and used our own codes to determine the $\moid$ and $\dtof$.  We found
differences in the MOID typically of less than $10^{-6}$~AU
($\sim$~100~km) and in $\dtof$ of less than 10 minutes. These results
are consistent across the cases discussed in this work, perhaps not
surprising as the encounters are all between orbits of modest
eccentricity and inclinations. We will discuss additional differences
further in cases (e.g. Xanthus in section~\ref{sec:xanthus}) where
they arise. The JPL integrator likely outperforms our own in precision
but the overall consistency of JPL computations with ours reinforces
our confidence in our results.

To determine appropriate conditions under which material accompanying
a catalogued asteroid might pose some hazard of collision with
Apophis, we use the fact that material released at low speed (which
includes most of the mass-shedding mechanisms listed earlier in
Section~\ref{sec:intro}) will follow its parent's orbit, with some
motion away from it but dispersing primarily forward or backward along
the orbit due to Keplerian shear. The orbital period of the bodies in
our sample are about 1 year, and material released at a few~ m/s will,
to an order of magnitude, move $\sim 10^5$~km before its orbit closes
on itself. During a single orbit, the material will acquire a time
offset $\sim 1$~hour from its parent, and this offset accumulates over
time. The chaotic e-folding time for these types of Earth-crossing
orbits is typically 100~yrs \citep{whi95}, and so we might expect any
material released by an asteroid to have been scattered from its orbit
on that time scale. From this, we determine that any material released
by an asteroid might be expected very roughly to have a $\moid \las
10^5$~km with Apophis and a $\dtof \las 100$~hours.

We note that a $\dtof \las 100$ hours corresponds to a much larger
distance than the $\moid \las 10^5$~km criteria: an asteroid
travelling at 30~km/s will travel about 10 million km in 100
hours. However, a larger along-track distance is reasonable because
the uncertainties in the asteroid positions are primarily along the
orbit due to Keplerian shear. As a result, material released from an
asteroid may end up dispersed over a very large volume, much much
larger than the volume of Apophis itself. Thus the chances of a
collision with asteroid-released material remain very low, though not
always negligible. Comets release material that disperses over an even
larger volume, but that may still create substantial flux, for
example as seen at Earth during a meteor shower. Nonetheless, the
risk of collision between Apophis and any of the asteroids discussed
in this work (or with material released from them) remains
exceptionally low.

We tighten the criteria slightly to reduce our list to the most
interesting objects. From our simulations of the 376 JPL objects and
396 NeoDys objects, we extract all those for which at least 1 clone, at
some point during the simulation, meets the condition
\begin{equation}
  \moid < 10^4~{\rm km~\hspace*{1cm} and \hspace*{1cm}}|\dtof| < 12~{\rm hours} \label{eq:closepass}
\end{equation}
which we will term an ``encounter''.  Since each of our 1000 clones at
this stage corresponds to an equally likely case within the formal
error distribution of the asteroid's orbital solution, this
corresponds to a $0.1$~\% chance of the asteroid in question passing
within 10,000~km of Apophis within $\pm 12$~hours of Apophis doing the
same. Fifteen asteroids meet this criterion and they will be discussed
in more detail below.

\section{Results and discussion} 

We will group our results into the following categories
\begin{enumerate}
\item Both JPL and NeoDys agree that there is a substantial ($2-100$\%) probability
  of an asteroid passing Apophis at a $\moid < 10^4$~km and a $\dtof < 0.5$~days
\item At least one of JPL or NeoDys indicate a marginal ($0.1-2$~\%) probability
  of an asteroid passing Apophis at a $\moid < 10^4$~km and a $\dtof < 0.5$~days
\item JPL and NeoDys disagree substantially. This is of some concern as it indicates
  that we do not have a consensus as to exactly how close Apophis will pass to the
  asteroid in question.
\item Other cases not strictly falling within our threshold but that are otherwise unusual.
\end{enumerate}  
Where orbital or physical parameters are discussed in the sections below, they will
be derived from those reported by the JPL Small Bodies Database unless otherwise
specified. The cases of interest are summarized in Table~\ref{tab:candidates}.

\subsection{Substantial probability of a close approach}

\subsubsection{4544 Xanthus}  \label{sec:xanthus}

Unique among our results is a close approach between Apophis and
asteroid 4544 Xanthus on 25 December 2026.  All of the 1000 clones of
the JPL and NeoDys solutions meet our threshold of $\moid < 10^4$~km
and $\dtof < 12$~hours, indicating that an encounter between these two
asteroids is certain.

4544 Xanthus has an absolute magnitude of $H=17.4$ \citep{MPO745323}
and a 1.3 km diameter. This considerably larger than Apophis, about 25
times more massive, assuming a similar density and an albedo of 0.15
\citep{chaharbin94}. Its JPL orbit is based on over 1800 observations
over 34 years, together with a single Doppler radar observation taken
from Arecibo soon after its discovery in 1990 \citep{ostcamcha91}. As
a result, it has an orbit code of 0, indicating that its orbit is very
well-determined.

\begin{figure}
  \centerline{\includegraphics[width=4in]{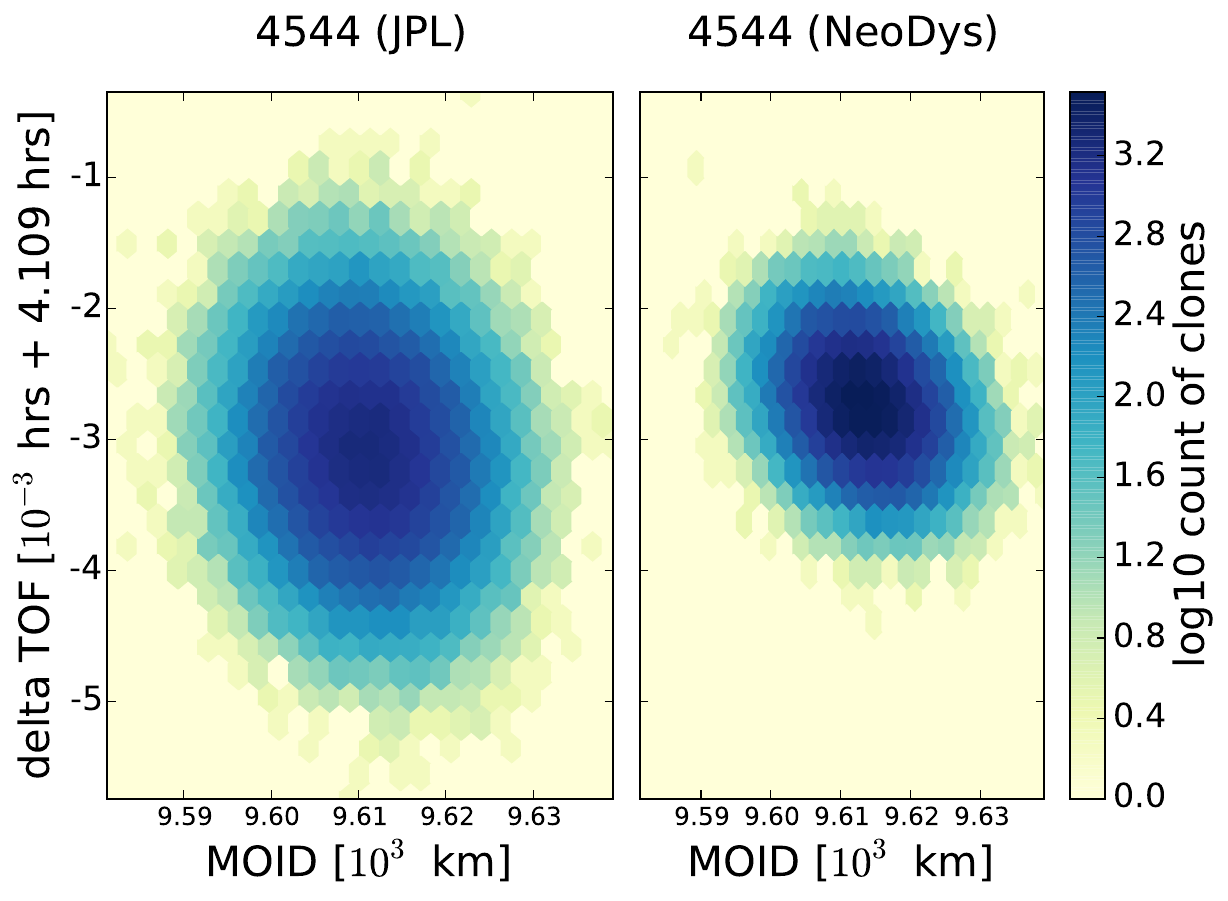}}
\caption{ The left panel shows the $\moid - \dtof$ for asteroid 4544 Xanthus with 99942
  Apophis for $10^5$~clones\label{fig:4544-moid-tof}. The density indicates
  the relative probability under the observational uncertainty, as determined
  by JPL (left) and NeoDys (right).}
\end{figure}

\begin{figure}
  \centerline{\includegraphics[width=3in]{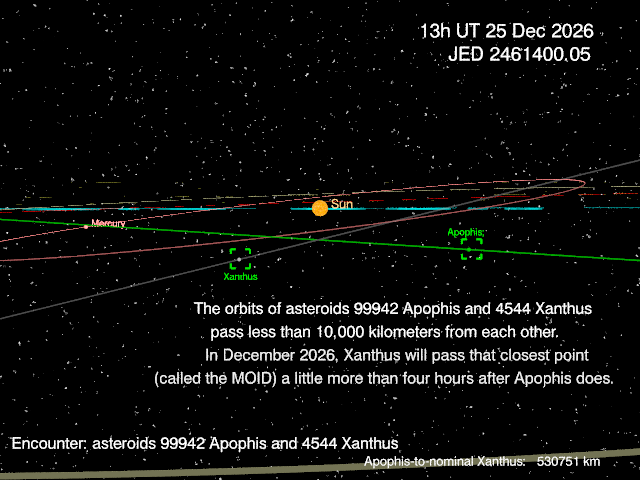}
              \includegraphics[width=3in]{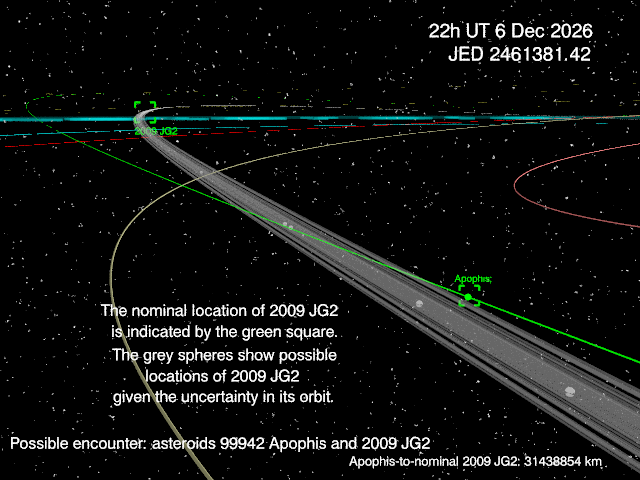}}
  \centerline{\includegraphics[width=3in]{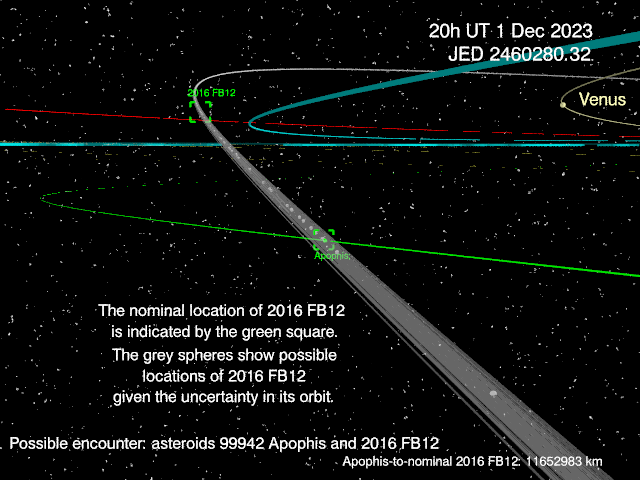}
              \includegraphics[width=3in]{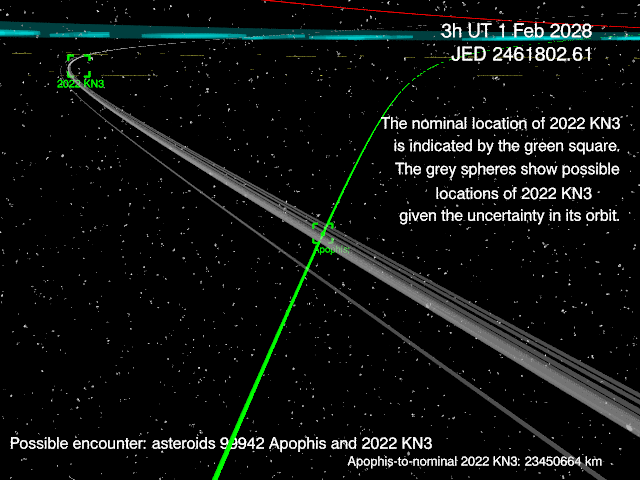}}
  \caption{ Illustrations of the Solar System context and encounter circumstances for 4544 Xanthus, 2009 JG2, 2016 FB12 and 2022 KN3. An animation is available in the HTML version of the paper for each panel. Each shows first a broad view of the Solar System, and then zooms in to illustrate the encounter geometry. The time and nominal asteroid-Apophis distance are also indicated. Duration of each animation is approximately 30 seconds. For this preprint, a compilation of these four animations is on YouTube at \url{https://youtu.be/k-1evaBWX60} \label{fig:animations}}
\end{figure}

In late 2026, Xanthus has mean MOID with Apophis of 9604 $\pm$ 6 km,
and a mean $\dtof$ of -4.1121 $\pm$ 0.0006 hours (JPL). From the
NeoDys data, Xanthus has mean MOID with Apophis of $9607\pm 6$ km, and
a mean $\dtof$ of $-4.1117 \pm 0.0004$ hours
(Figure~\ref{fig:4544-moid-tof}).  Xanthus will pass its mutual
$\moid$ with Apophis just over four hours after Apophis itself passes
that point: see Figure~\ref{fig:animations} for an animated
illustration of the Solar System context and encounter circumstances.

The fact that the JPL and NeoDys distributions are not identical in
Fig~\ref{fig:4544-moid-tof} is a result of their independent
operations. The JPL and NeoDys orbits are each computed by a group of
expert astronomers on the basis of the same data using similar
techniques. However, the data invariably contains some measurement
uncertainty. Each group compensates for this using their own
weightings and corrections, based on their respective extensive
collective experience, so the final results may differ. This
independence of one group from the other allows extensive
cross-checking and verification, and increases the robustness of our
near-Earth asteroid catalogues.

Because of the close agreement between JPL and NeoDys in
Fig~\ref{fig:4544-moid-tof}, with each heavily overlapping within the
other's formal errors, we conclude with high confidence that
Xanthus has a MOID with Apophis of less than 10,000 km and will
arrive at that point just over 4 hours after it in Dec 2026. Despite
their relatively small MOID, the 4 hour delay means that they are in
no danger of collision. In terms of physical distance, Xanthus reaches
a minimum distance of about 528,000 km with Apophis on JD 2461400.1
(25 Dec 2026) as they pass each other at a relative velocity of 11
km/s.

The encounter of Xanthus with Apophis is one where our results differ
more than usual with JPL Horizons, though only slightly. We find
differences in the MOID of $\approx300$~km between our and JPL's
propagation of the orbits, and a difference in the $\dtof$ of 1.3
minutes. This is larger than the spread in the $\moid$ and $\dtof$
seen in Fig~~\ref{fig:4544-moid-tof} , which are only $\approx 10$~km
and a few seconds respectively. Though we differ in detail with the
JPL results, we do so only at the few parts per million level in
overall position ($300~\rm{km} \approx 2 \times 10^{-6}$~AU), and our
MOIDs and $\dtof$ differ only at the $(300~\rm{km}/9600~\rm{km})
\approx 3$\% and $1.3~\rm{min}/(4.1~\rm{h} \times 60~\rm{min}) =
0.5$~\% levels. The difference likely stems from the frequent
encounters between Xanthus and the Earth: the JPL Small-Body Database
records three of them in 2021-2022, and four of them in
2023-2024. These encounters occur at modest distances (typically
0.2~AU) and at low relative speed (10~km/s), and tax attempts to model
Xanthus' motion accurately. Nonetheless, our results are consistent
with those of JPL Horizons, and the slight discrepancies arise in this
case from an asteroid which is particularly difficult to model.

It is extremely unlikely that the orbital parameters as computed by
JPL or NeoDys could harbour uncertainties large enough to permit a
collision. Even if one or more of the observations upon which those
orbital parameters is based is of poor quality or even outright wrong,
the overall computation is tightly constrained by the many other
observations and the laws of motion. The one observation that perhaps
deserves a second look is the radar Doppler observation
\citep{ostcamcha91}, which provides a single data point which strongly
affects the final orbit determination. Though there is no reason to
expect that reviewing the radar data will improve its precision or
accuracy, we recommend that this object, along with some more
problematic cases we discuss later, have their orbit computations
revisited on a purely precautionary basis.

Of more concern than direct collision is the question of whether
Apophis could collide with material accompanying Xanthus. This could
result in a perturbation of its future path that could affect its
impact probability with Earth. There are a number of possibilities.

\begin{itemize}
\item {\bf Satellite asteroids:} There is no evidence to suggest that
  Xanthus is a multiple asteroid but it may be. Stable satellites of
  Xanthus could only reside within its Hill sphere. For a 1.3~km
  diameter asteroid with a density of 2000~kg/m$^3$ this translates to
  about 200~km, and in practice most NEA multiples remain at much
  smaller distances from each other. For example, for the 87
  near-Earth and Mars-crossing asteroids with satellites listed in
  \cite{joh19}, the median separation was 3.3 km and the largest was
  378~km. Given that Apophis and Xanthus will pass each other at a
  much larger distance ($> 500,000$~km), it is unlikely that Apophis
  could collide with satellites of Xanthus, if any exist.

\item {\bf Cometary activity:}  We are not aware of any reports of
  cometary activity from this asteroid, and its Tisserand parameter
  with respect to Jupiter is 5.834, more typical of asteroids than
  comets. It is unlikely that it has produced a traditional cometary
  meteoroid stream.

\item {\bf Active asteroid:} Xanthus could have low-level undetected
  mass loss, which would make it an ``active asteroid''. There are a
  handful of active asteroids known \citep{jewhsiaga15} which may
  release material through a wide array of mechanisms.  Material
  travelling with Xanthus will be particularly difficult to detect
  with telescopes on Earth if it is low in dust production, dust
  having a large optical cross-section for its mass. For example, the
  OSIRIS-Rex spacecraft revealed that asteroid 101955 Bennu was
  surrounded by 1-10~cm particles ejected from its surface at speeds
  of a few m/s \citep{hermalnol19, laudelben19}. These particles are
  thought to be ejecta due to meteor impacts, but no dust was
  reported. Particles were seen to persist for several days in some
  cases, and some were on trajectories that would allow them escape
  Bennu's gravity entirely. It is possible that Xanthus is producing
  material in this way. The material discovered near Bennu was a
  surprise, and other mechanisms could also produce unexpected
  material travelling with almost any asteroid.

\item {\bf Gravitational impulse:} To be of concern, any change in
  Apophis' velocity would have to be able to move Apophis' trajectory
  substantially relative to the locations of various ``keyholes''
  \citep{valmilgro03,farchecho13} which could lead to future impacts with our
  planet. This would require moving Apophis at least tens of km over the course
  of several years, or an impulse of order $10^{-3}$m/s.

Assuming a density $\rho=2000$~kg~m$^{-3}$ for Xanthus, the expected
$\Delta v$ to Apophis for the Dec 2026 encounter is (from
Eq~\ref{eq:impulse_approx})
\begin{equation}
\Delta v \approx 10^{-10}   \left( \frac{\rho}{2000~\rm{kg/m}^3} \right) \left( \frac{11~\rm{km/s}}{v_{rel}} \right) \left( \frac{5.3 \times 10^5\rm{km}}{d} \right)~\rm {m/s}
\end{equation}  
and is entirely negligible. To provide some context, the largest
gravitational impulse that Xanthus could exert on Apophis without a
collision --- that is, assuming they passed each other as closely as
possible ($d \approx 1$~km) --- is $\sim 10^{-4}$~m/s. So even a very
close pass by a kilometre-class asteroid could perturb Apophis'
orbit only marginally via purely gravitational effects.

\end{itemize}

If particles were being ejected from Xanthus by any mechanism, it is
possible that Apophis could collide with this debris as the asteroids
pass each other. Debris from Xanthus (if any) could pass closer to
Apophis than Xanthus itself, and could strike it. The impact of dust
particles in the sub-mm range is unlikely to have a substantial effect
\citep{wie15} on Apophis' path, but the impact on Apophis of even a
single 10~cm particle such as seen at Bennu at a relative speed of
11~km/s and a typical asteroid density of 2000~kg~m$^{-3}$
\citep{briyeohou02} would release about 20 MJ of kinetic energy, the
equivalent of 20 sticks of
dynamite. 
Such an event could have a non-negligible effect on the impact
probability of Apophis with Earth at a later date.

The presence or absence of macroscopic material accompanying Xanthus
cannot easily be determined by current telescopes, particularly if the
mass-shedding mechanism doesn't produce much dust, as was the case for
particles seen by OSIRIS-Rex at Bennu. Not all asteroids visited by
spacecraft have had such particles reported, and even if such material
exists, the probability of a collision with Apophis is low. However,
we suggest telescopic monitoring of the encounter between Apophis and
Xanthus to assess whether such an impact occurs, for example, by
looking for the dust production that could be expected from a
hypervelocity impact.

As seen from the Earth in late Dec 2026, the two asteroids will
approach each other closely on the sky (an 'appulse'), According to
JPL
Horizons\footnote{\url{https://ssd.jpl.nasa.gov/horizons/app.html\#/}},
the event will be difficult to observe from Earth as it occurs at a
solar elongation of only 47 degrees in the evening sky, in the
constellation Capricorn. Apophis will be at $m_v$ = 20.7 and Xanthus at
$m_v$ = 19.2, accessible in principle with professional calibre
telescopes but interference by scattered sunlight could be an issue
for ground-based observers. An animated illustration of the event is
in Figure~\ref{fig:animations}.

Out of curiosity we note that the two asteroids will not reach
naked-eye visibility to each other during this close approach, though
each would be easily visible from the other through binoculars.
Xanthus will reach a peak visual magnitude $m_v = 8.1$ as seen from
Apophis, while an observer on Xanthus will see Apophis brighten to
$m_v = 9.2$.  For reference, Apophis would reach naked-eye visibility
($m_v=6$) from any particular asteroid (assuming a heliocentric
distance of 1 AU and zero phase) at a mutual distance of 0.0024~AU or
358,800~km. Asteroid 2014~AD16 is the only object in this study for
which it can be confidently said that it will approach Apophis this
closely, though phase effects keep the apparent magnitude of Apophis
as seen by an observer on that asteroid below naked eye visibility
(see Section~\ref{sec:2014AD16}).

The encounter between Apophis and Xanthus is unique over the next few
years, but there are other cases of modest probability of close
encounters between Apophis and other asteroids. We will first discuss
the chance of a direct collision in each case (which will turn out to
be nil) and then discuss the possibility of a collision material
accompanying these asteroids in aggregate at the end.

\subsubsection{Asteroid 2009 JG2}

\begin{figure}
  \centerline{
    \includegraphics[width=3in]{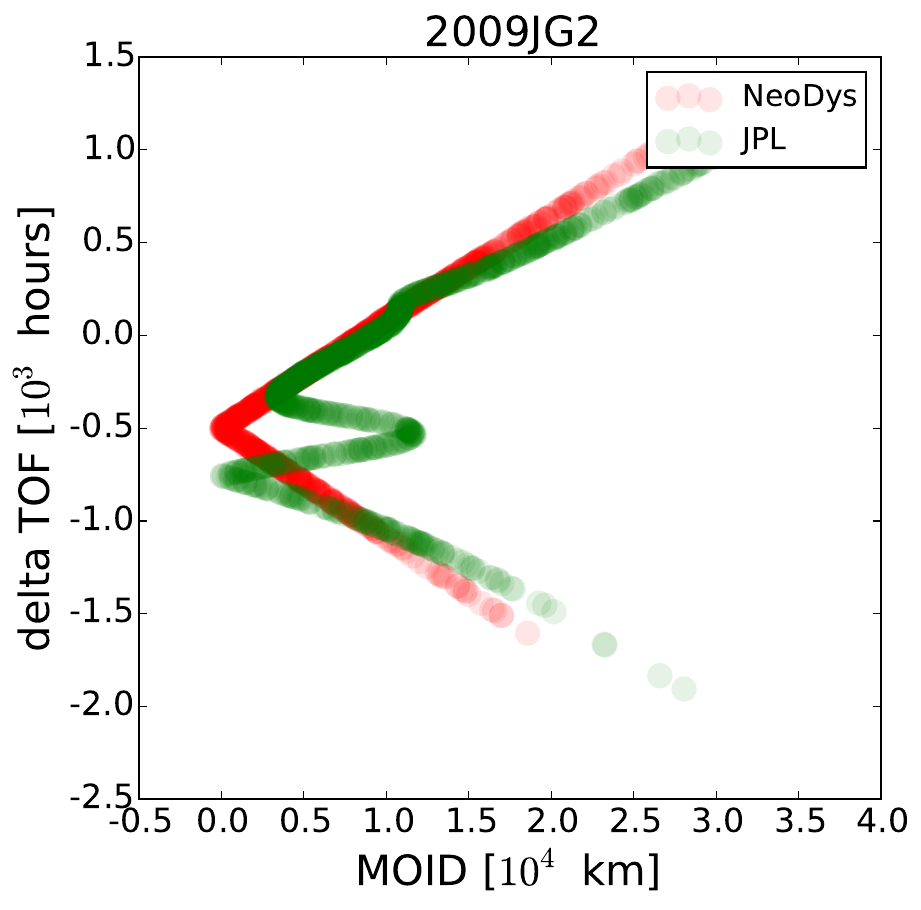}
    \includegraphics[width=3in]{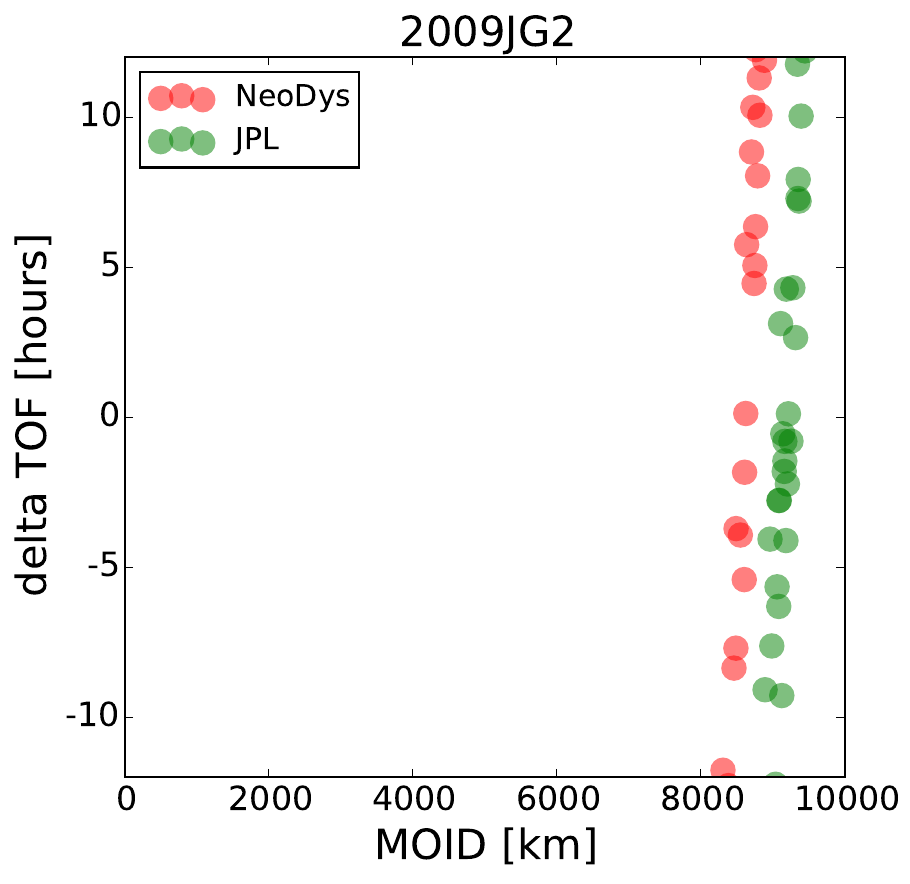}
  }
\caption{The $\moid - \dtof$ for the December 2026 approach of asteroid
  2009 JG2 to 99942 Apophis ($10^3$~clones). The right panel is
  zoomed in on a region near the origin. \label{fig:2009JG2-moid-tof}}
\end{figure}

Asteroid 2009 JG2 is a 100~m diameter asteroid that was only observed
briefly (for 23 days) in 2009 and it has an orbit code of 8 (poor).
Its $\moid-\dtof$ plot is presented for the encounter which occurs in
late 2026 in Fig~\ref{fig:2009JG2-moid-tof}. The uncertainty in its
orbit determination is reflected primarily in its time of arrival at
the MOID with Apophis, which is uncertain by several weeks in either
direction. Both JPL and NeoDys orbital solutions yield a 2\% chance of
having an encounter (as defined by Eq.~\ref{eq:closepass}) with
Apophis.

The distribution of the NeoDys clones in
Fig~\ref{fig:2009JG2-moid-tof} is rather linear (in fact they trace
out an elongated ellipse, reflected across the $y$-axis at its
crossing) while the JPL clones have their basic elliptical shape
distorted. The distortion of the JPL clone cloud arises due to a
moderately close encounter of 2009~JG2 with Venus in early 2021.
Despite the differences between the JPL and NeoDys solutions, by
examining the region nearest the origin in
Fig~\ref{fig:2009JG2-moid-tof}, we can see that both agree that this
asteroid will not collide with Apophis. Even if the two asteroids do
both happen to be at the MOID at the same time ($\dtof=0$) they will
still remain 8000-9000~km apart. It is also possible that Apophis and
2009~JG2 have intersecting orbits ($\moid = 0$) but in this case,
2009~JG2 will pass that intersection point 20-30 days after
Apophis. An illustration of the event is presented in
Figure~\ref{fig:animations}.

While the uncertainty in the currently known orbit of 2009~JG2 allows the
possibility of a zero $\moid$ or of a zero $\dtof$, the case where
both $\moid$ and $\dtof$ are zero is excluded. Both JPL and NeoDys
solutions agree that the asteroids will not collide.

To monitor whether Apophis collides with any material on the orbit of
2009 JG2, observations should be taken around the time when Apophis is
crossing the orbit of the asteroid, but the time of the possible
closest approach is spread over tens of days. Apophis is at its
$\moid$ with the nominal orbit of 2009~JG2 on 6 Dec 2026 at a solar
elongation of only 41 degrees, in the constellation
Sagittarius. Apophis will be at $m_v = 20.7$; the nominal location of
2009 JG2 will be several degrees away and closer to the Sun.

\subsubsection{Asteroid 2016 FB12}

\begin{figure}
  \centerline{
    \includegraphics[width=3in]{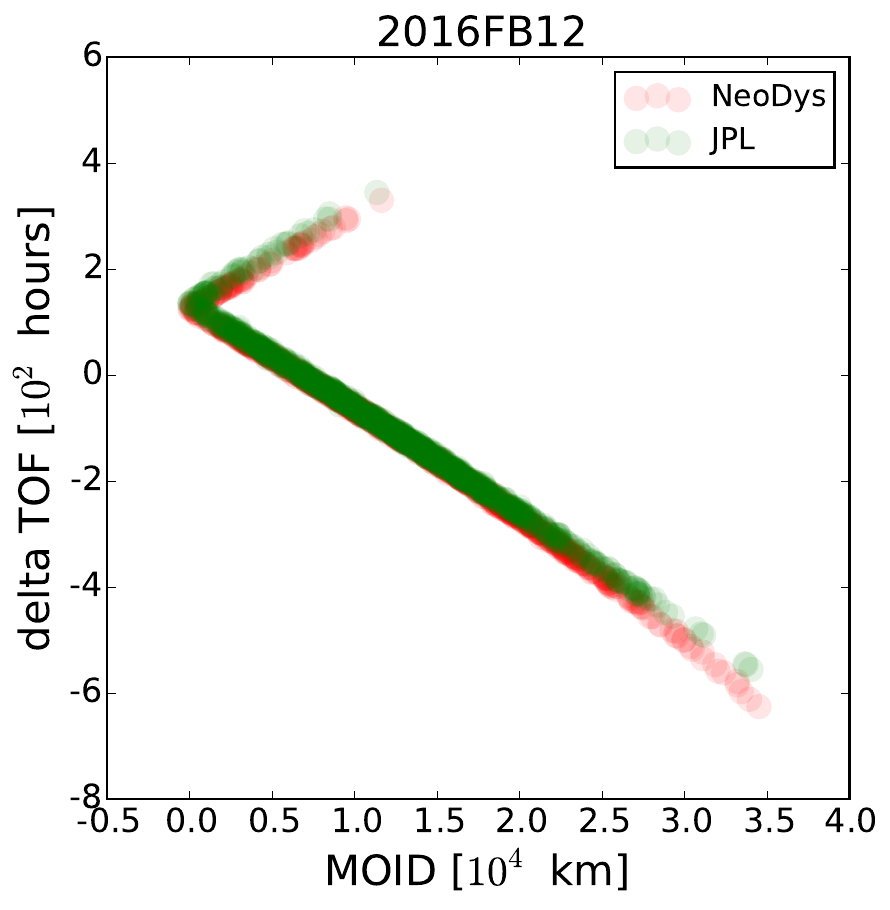}
    \includegraphics[width=3in]{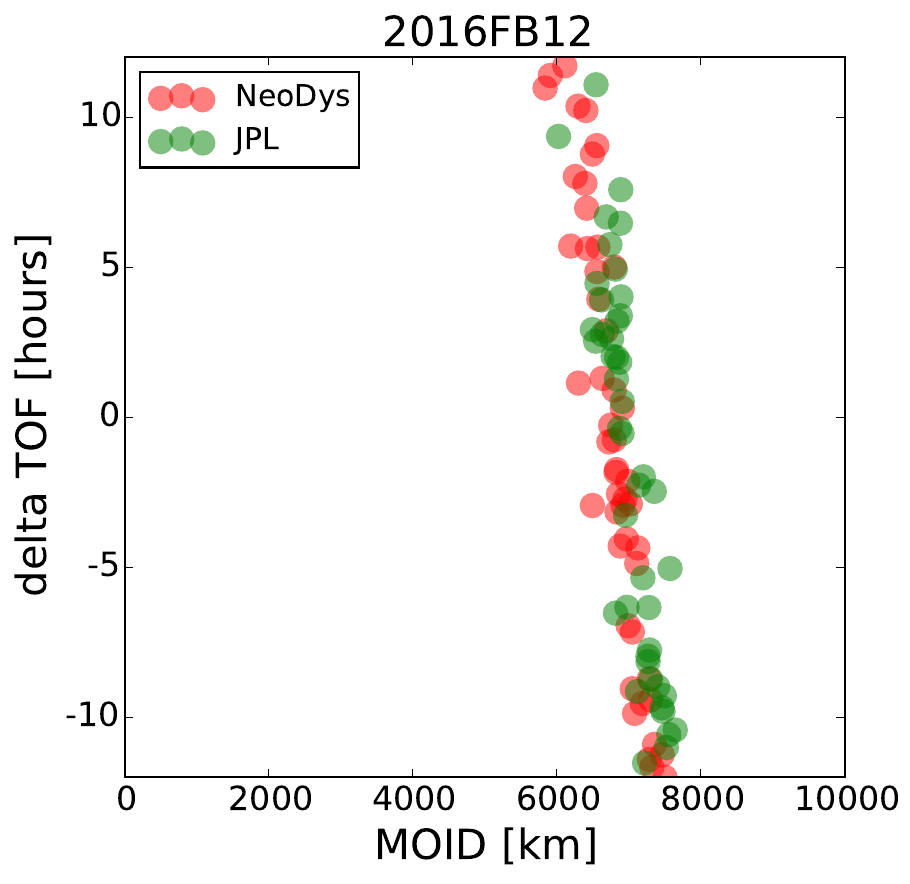}
  }
\caption{The $\moid - \dtof$ for the December 2023 approach of asteroid
  2016 FB12 to 99942 Apophis ($10^3$~clones). The right panel is
  zoomed in on a region near the
  origin. \label{fig:2016FB12-moid-tof}}
\end{figure}

Both JPL and NeoDys predict a $\approx 2$\% chance that 2016 FB12 will
have an encounter with Apophis in early Dec 2023.
Figure~\ref{fig:2016FB12-moid-tof} shows the distribution of $\moid$
and $\dtof$. The distributions are similar for the JPL and NeoDys
solutions, though not identical.

2016~FB12 was only observed briefly (for 4 days) in 2016 and
it has an orbit code of 8 indicating its path is rather poorly known,
as a direct result of the shortage of observations. Its absolute magnitude
of 22.6 means its diameter is roughly 20~m. 

For 2016~FB12, Figure~\ref{fig:2016FB12-moid-tof} also shows (on the
right) the region nearest the origin.  We can see that, despite the
uncertainty in the precise position of this asteroid, both the JPL and
NeoDys solutions indicate that the asteroid will not collide with
Apophis, because the MOIDs are well constrained away from zero. Though
these solutions do allow that both asteroids could be at the MOID at
the same time, both would still be 6000-7000~km apart should this
occur. And the orbital solution allows that the $\moid$ could be zero,
but in that case 2016~FB12 will arrive at the MOID days after
Apophis. The $\moid = \dtof = 0$ (collision) case is not consistent
with the observations of the asteroid's orbit to date. The chance of a
direct collision is nil.

If one wished to observe Apophis at its Dec 2023 encounter to monitor
for possible collisions with accompanying material, the time of close
approach is uncertain by tens of days, and the circumstances
unfavourable.  Apophis will be at a solar elongation of only 33
degrees at $m_v = 21.6$ in Virgo.  The nominal location of 2016~FB12
is nearby in the same constellation, but its small size means its
apparent magnitude will be fainter than $28$. The encounter
circumstances are illustrated in Figure~\ref{fig:animations}.

\subsubsection{Asteroid 2022~KN3}
\begin{figure}
  \centerline{
    \includegraphics[width=3in]{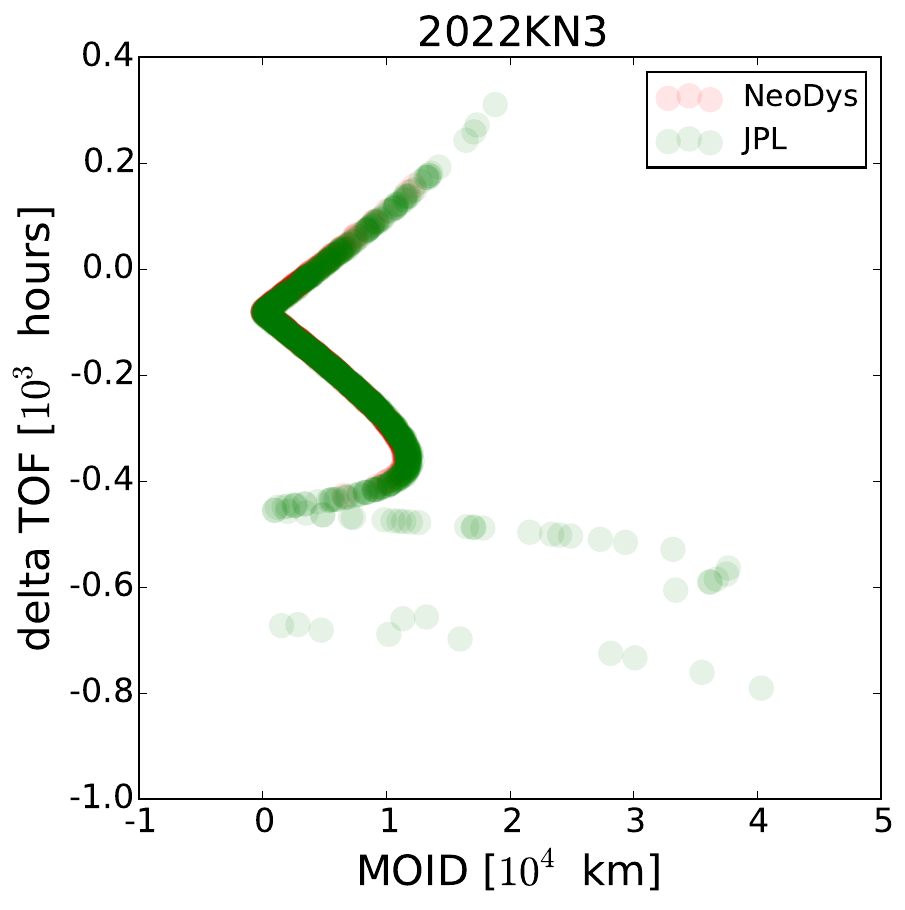}
    \includegraphics[width=3in]{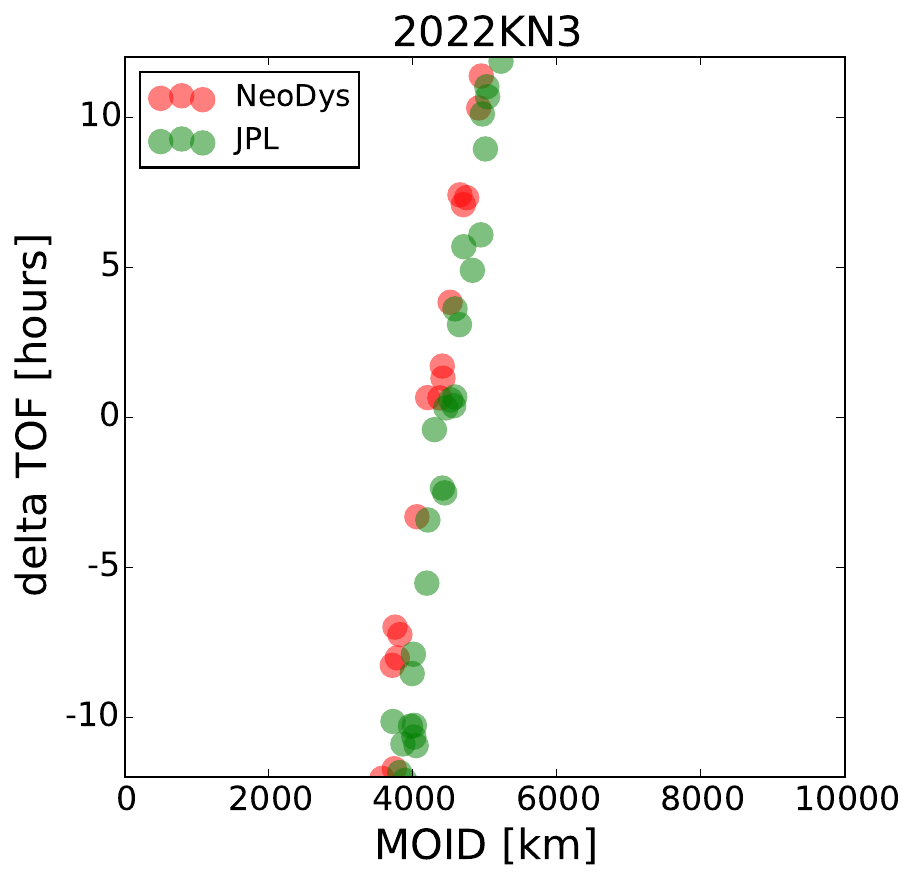}
  }
\caption{The $\moid - \dtof$ for the February 2028 approach of
  asteroid 2022~KN3 to 99942 Apophis ($10^3$~clones). The right
  panel is zoomed in on a region near the
  origin. \label{fig:2022KN3-moid-tof}}
\end{figure}

Asteroid 2022~KN3 was observed for 12 days only, and has an orbit code
of 8 (poor).  Its absolute magnitude of 24.8 indicates an approximate
diameter of 40~m.  The $\moid - \dtof$ plot of asteroid 2022~KN3 is
shown in Figure~\ref{fig:2022KN3-moid-tof} for its encounter with
Apophis in early 2028.

The distribution shows some distortion, in this case owing to close
approaches with Earth in May 2025, and in May 2028 (the asteroid has a
semimajor axis near the 3:1 mean-motion resonance with Earth). But by
examining the plot's inner region more closely, both JPL and NeoDys
predict that Apophis and 2022~KN32 will pass each other at a distance
of 4000 km at most, even if both are at the MOID at the same
time. Both JPL and NeoDys agree there is no chance of a direct
collision, and any impacts associated with this encounter would be due
to material (if any) following the orbit of 2022~KN3.

If any collision with debris will occur, it is most likely as Apophis
passes its MOID with 2022 KN3. Apophis crosses its $\moid$ with the
nominal orbits of this asteroid in 2028 at a solar elongation of 89
degrees in the constellation Cetus. These relatively favourable
circumstances have Apophis at $m_v = 19.3$, though 2022~KN3 will be
very faint at 25.8, and nominally a few degrees away on the sky. An
illustration of the encounter circumstances is in
Figure~\ref{fig:animations}.

\subsubsection{Collision with material associated with these asteroids}

The possibility of collision with material accompanying any of the
last three asteroids discussed is comparable in many ways to that of
Xanthus.  However, the latter three are all much smaller (20-100~m
versus 1300~m for Xanthus), reducing the likelihood of stable
satellites and the amount of debris that might be in their
neighbourhood.  Nevertheless, we would encourage observations of
Apophis near their times of close approach to confirm or deny the
occurrence of impacts by debris. We suggest observers monitor Apophis,
who's on-sky position is well-known, rather than attempting to recover
the asteroids themselves as their ephemeris uncertainties are much
larger. If a collision occurs, the resulting dust will be revealed by
the ejection of dust from Apophis, even if the colliding asteroid
itself goes unseen.

Even before the times of encounter, additional observations of these
asteroids would be useful in further constraining their orbits;
however, ephemeris uncertainties make this difficult. Regardless,
these objects' orbits could be improved by 1) re-measurement or
re-analysis of existing observations and/or 2) the identification of
additional precovery observations in image archives.

Though JPL and NeoDys agree that collisions will not occur, we will
see in the next section that discrepancies between the JPL and NeoDys
predictions can occur for poorly known orbits, and there remains an
element of unreported uncertainty that is associated with these cases,
which we turn to now.

\subsection{Conflicting case: 2016 CL18} \label{sec:conflict}

We discuss here the one asteroid we examined whose orbital encounter
circumstances with Apophis differ significantly between JPL and NeoDys
solutions. Despite both sources confidently predicting non-collision,
the divergence between their results undermines our ability to assert
this outcome with absolute certainty. This is an object that would
benefit from additional observations and analysis.

\begin{figure}
  \centerline{
    \includegraphics[width=5in]{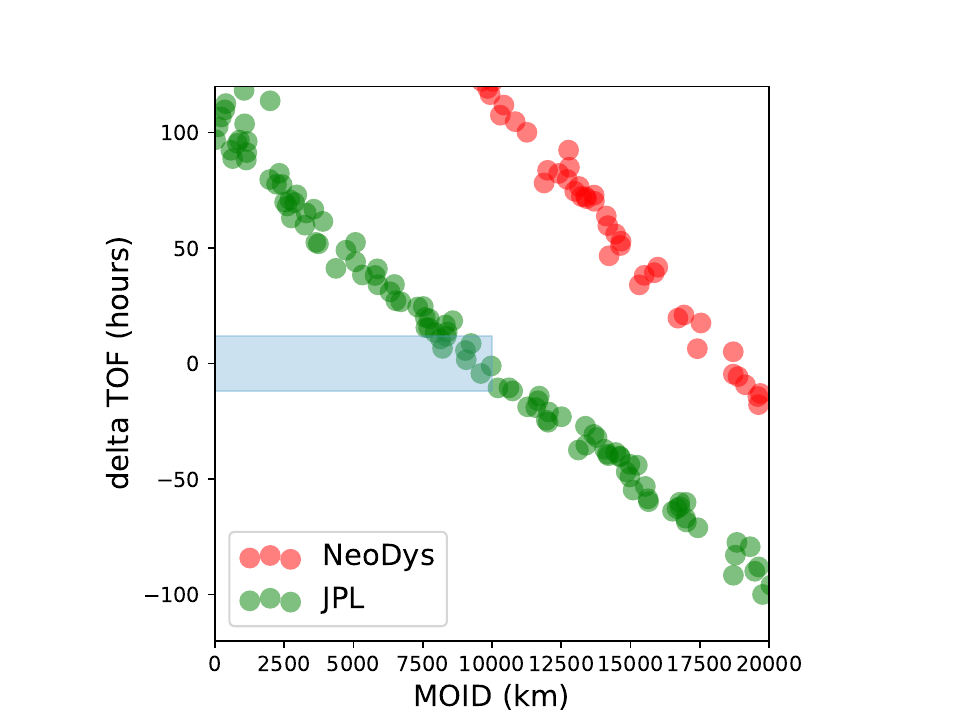}
  }
\caption{The $\moid - \dtof$ plot for the December 2026 approach of
  asteroid 2016~CL18 to 99942 Apophis ($10^3$ clones). The threshold
  region of Eq~\ref{eq:closepass} is shown in light
  blue. The JPL and NeoDys solutions put the asteroid into non-overlapping regions of space approximately 10,000~km apart. \label{fig:2016CL18-moid-tof}}
\end{figure}

Asteroid 2016 CL18 is a 50-meter Apollo class asteroid that was only
observed for 7 days near its time of discovery. Its orbit is very
imperfectly known (orbit code 9). Its $\moid-\dtof$~plot is shown in Fig
~\ref{fig:2016CL18-moid-tof} for late 2026.

This asteroid passes our threshold criterion for an encounter
(Eq~\ref{eq:closepass}) for many of the clones in its JPL sample, but
none in the NeoDys sample.  Though it is perhaps not surprising that
JPL and NeoDys should differ markedly given the sparseness of the data,
the technical point of concern is that both groups put the object with
high confidence into regions of space which hardly overlap. In this case,
we are forced to assume that the real uncertainty in asteroid's position
is of order the difference between the distributions predicted by these two
groups, tens of thousands of km in the $\moid$ and days in $\dtof$.

The solution to this problem is additional observations of these
objects, as we find that the JPL and NeoDys solutions almost always
agree when there is good data coverage.  Additional data could come
from new telescopic recovery observations as well as archival
precovery observations. Extending the observational arc of these
objects will allow more refined and confident statement to be made
about the chances of collision, and observations of Apophis near the
time of orbit crossing with 2016 CL18 can also determine whether or
not a collision has occurred. The observing circumstances for the 2016
CL18 encounter with Apophis are similar to those of Xanthus described
earlier, both encounters taking place in late December 2026.

\subsection{Low probability of an encounter}

The ``Low probability'' objects in Table~\ref{tab:candidates} are
those for which only one or two clones in either of the asteroid
catalogues matched the criteria of Eq~\ref{eq:closepass}.  These
universally have poorly known orbits and so the chance of collision is
hard to assess. We list them here because improvements in their
orbit determinations, from recovery or precovery observations or other
means would be useful, and would allow us to assess the probability of
a collision better. We will only discuss one in detail.

\subsubsection{2001 FB90}

Asteroid 2001 FB90 was observed for only 11 days (orbit code 9) making
it difficult to localize in its orbit. We mention it here because of
its large size ($H=19.8$ or 380~m) make it the third largest object in
Table~\ref{tab:candidates}, and because it is the only object on that
table with a Tisserand parameter with respect to Jupiter ($T_J =
2.976$) in the cometary regime. Both of these increase the probability
of material following along with 2001~FB90 and posing a risk of
collision with Apophis.  The approach takes place in Sep 2028 and
will be unobservable as Apophis will be less than 10 degrees from the
Sun on the sky. If an impact occurs, the effects (e.g. dust
production) may be visible when Apophis moves further away from the
Sun in late 2028.

\subsection{Other cases of note}

We include in our discussion a few cases which are not strictly within
our chosen threshold for an encounter but are otherwise of interest.

\subsubsection{373135}

\begin{figure}
  \centerline{
    \includegraphics[width=4in]{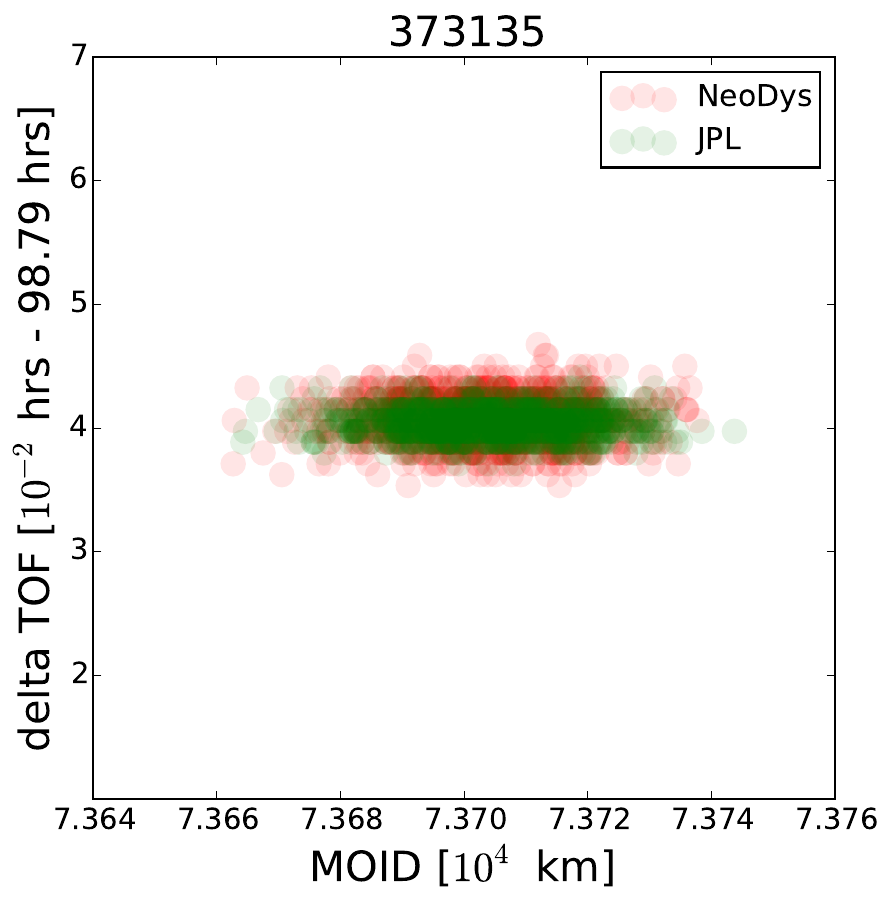}
  }
\caption{The $\moid - \dtof$ for the December 2023 approach of
  asteroid 373135 to 99942 Apophis ($10^3$
  clones). \label{fig:373135-moid-tof}}
\end{figure}

Asteroid 373135 (orbit code 0) has a MOID of $73700 \pm 13$~km and a
$\dtof$ of $4.11793 \pm 0.00005$~days
(Figure~\ref{fig:373135-moid-tof}) during its well-localized encounter
with Apophis in late 2023. It is mentioned because of its relatively
large size of $1050\pm400$~m \citep{nugmaibau16} which may increase
the likelihood of material in its vicinity. The diameter measurement
is from NEOWISE and indicated a low (0.025) albedo. If the low albedo
is taken as a sign of a possible carbonaceous composition, 373135 may
resemble 101599 Bennu, which is known to have 1-10~cm particles in its
vicinity \citep{hermalnol19, laudelben19}. Apophis passes its MOID
with 373135 while within 30 degrees of the Sun and will not re-emerge
to easy visibility for some months later.

\subsection{2014 AD16} \label{sec:2014AD16}

\begin{figure}
  \centerline{
    \includegraphics[width=4in]{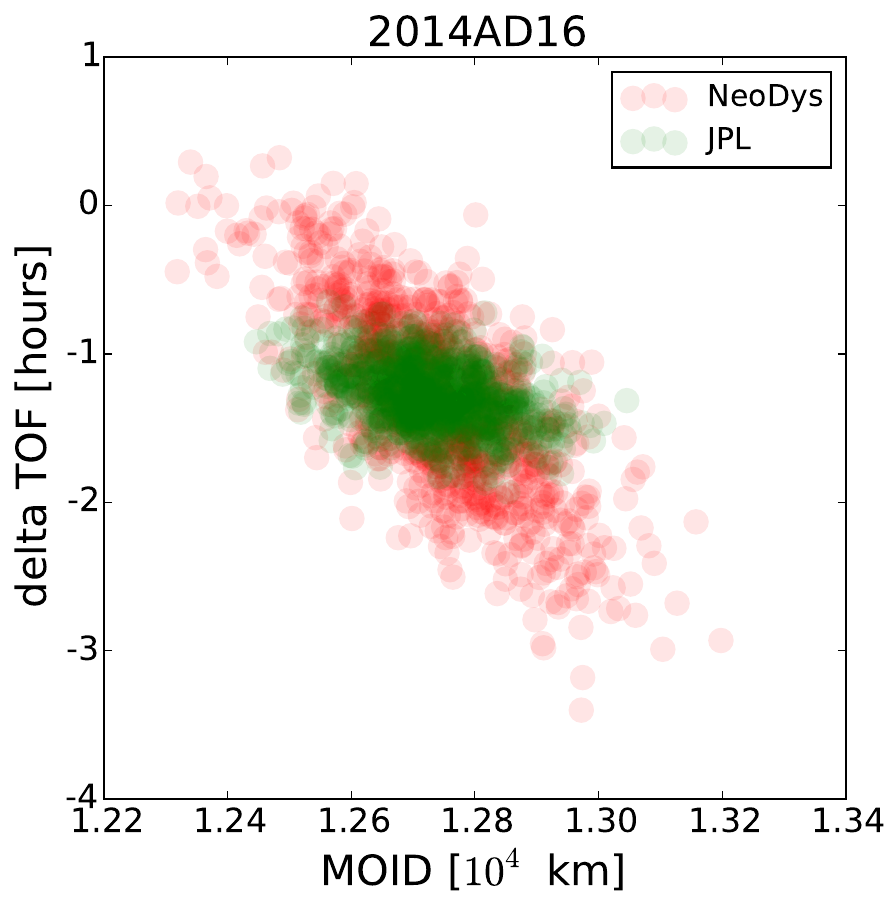}
  }
\caption{The $\moid - \dtof$ for the July 2017 approach of asteroid
  2014~AD16 to 99942 Apophis ($10^3$~clones). \label{fig:2014AD16-moid-tof}}
\end{figure}

Asteroid 2014 AD16 was observed only briefly (3 day arc). However it
was observed by radar and its orbit is consequently quite well known
(orbit code 4). It is included here because, during its 2027 encounter
with Apophis, though its $\moid$ (12700~km) is just outside our
threshold, its $|\dtof|$ at just over 1~hour is among the shortest
computed in this work (Figure~\ref{fig:2014AD16-moid-tof}). The
asteroids will approach to within 133,000~km of each other in July
2027. To an observer on 2014~AD16, Apophis will reach a visual
magnitude of 6.9 at a phase of 75~$\deg$ during this approach;
2014~AD16 will have $m_v = 14$ when viewed from Apophis at this time.

Though the formal chance of a collision with Apophis is zero,
2014~AD16's orbit determination is highly dependent on a small number
of radar observations, and might benefit from a re-analysis. 2014~AD16
is a small object ($H=27.4$, 10~m class) and so is unlikely to have
substantial amounts of material around it. Apophis will be at its MOID
with 2014~AD16 under difficult observing circumstances from Earth,
while 49~degrees from the Sun in Leo at $m_v = 21.3$.

\subsubsection{25143 Itokawa}

\begin{figure}
  \centerline{
    \includegraphics[width=4in]{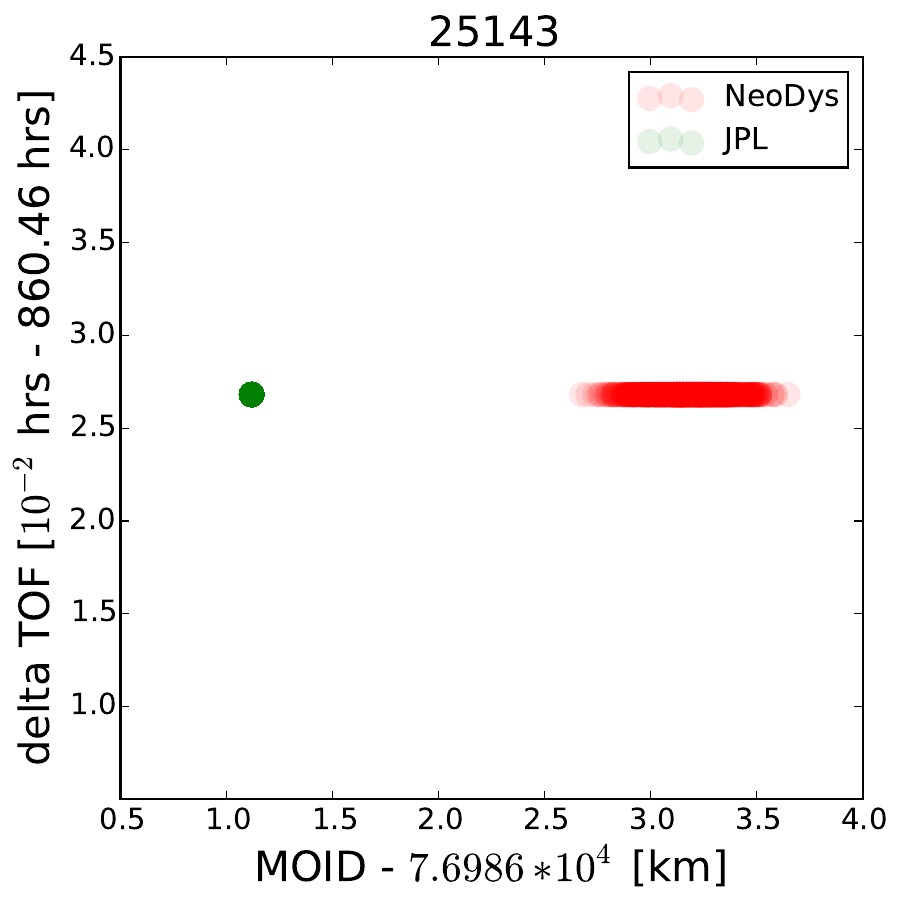}
  }
\caption{The $\moid - \dtof$ for the October 2025 approach of asteroid 25143 Itokawa to 99942 Apophis ($10^3$~clones). \label{fig:25143-moid-tof}}
\end{figure}

Asteroid 25143 Itokawa (orbit code 0) has a MOID with Apophis of
77,000 km and passes that MOID 35 days before Apophis in 2025
(Figure~\ref{fig:25143-moid-tof}). This event doesn't meet our usual
threshold and there is no chance of direct collision, but is mentioned
because Itokawa is known to have accompanying material, in the form of
the MINERVA lander released by JAXA's Hayabusa mission and which
is expected to have escaped Itokawa's gravity well.  MINERVA is
only 10~cm in size and 591~g \citep{yoskubnak06}.

Though the chance of MINERVA impacting Apophis is very low, such an
event given the relative speed of 6.8~km/s would release approximately
14 MJ of kinetic energy, equivalent to 3 kg of TNT. For reference, the
370 kg Deep Impact probe's 2005 collision with comet 9P/Tempel at a
relative speed of 10.2 km/s released 19 GJ of energy \citep{tayhan05};
the Double Asteroid Redirection Test (DART) 580~kg impactor released
11 GJ of energy in its 2022 collision at a relative speed of 6.1~km/s
\citep{dalernbar23}. The 2025 approach of Itokawa to Apophis will occur
while Apophis is 9 degrees from the Sun and Apophis will not re-emerge
into the night sky until some months later, making monitoring of this event
quite difficult.

\subsubsection{Comets}

Seven comets appear in the subsample we analyzed in
Section~\ref{sec:simulatedsample}:  C/1882 F1 (Wells), C/1917 F1
(Mellish), C/1917 H1 (Schaumasse), C/1961 T1 (Seki), C/1990 N1
(Tsuchiya-Kiuchi), C/2015 P3 (SWAN), C/2021 P4 (ATLAS). Each is
currently far from the inner solar system, and they pose no danger of
a direct collision with Apophis. However, these comets are the most
likely objects in our sample to have substantial material moving along
their orbits. Because these comets have large orbits, the density of
material along their orbits is on average low. The two with the
smallest orbits (Mellish:~$a \approx 27$~AU, Seki:~$a \approx 87$~AU)
are Halley-type comets. Comet Mellish has been associated with the
weak December Monocerotids meteor shower among others \citep{dru81,
  nesvauhaj16} so there is known to be some material along its
orbit. This work has not examined the possibility of Apophis
encountering material in the meteoroid stream of these comets in
detail; this possibility remains to be looked at in future work.

\begin{table}
  \centering
    \begin{tabular}{|c|c|c|c|c|c|c|c|c|c|c|}
        \hline
 Encounter  & \textbf{Name} & \textbf{a} & \textbf{e} & \textbf{i} & \textbf{T$_{Jup}$} & \textbf{H} & \textbf{Diam.} & \textbf{$v_{rel}$} & \textbf{Orbit}  & \textbf{Encounter}\\
  probability&  & \textbf{(AU)}& & \textbf{(deg)} & & & \textbf{(m)} & \textbf{(km/s)} & \textbf{code} & \textbf{date} \\ \hline

              & 99942 Apophis   & 0.923 & 0.191 & 3.34  & 6.464 & 19.1  & 340  &  - & 0 &   --- \\ \hline \hline
               & 4544 Xanthus   & 1.041 & 0.250 & 14.1  & 5.834 & 17.4  & 1300 & 11 & 0 &  25 Dec 2026 \\ \cline{2-11} 
 High to       & 2009 JG2       & 2.032 & 0.659 & 2.73  & 3.500 & 22.6  & 100  & 10 & 8 &  6 Dec 2026 \\ \cline{2-11}
 moderate      & 2016 FB12      & 1.332 & 0.264 & 7.73  & 4.872 & 26.0  & 20   & 11 & 8 &  1 Dec 2023 \\  \cline{2-11}
               & 2022 KN3       & 2.069 & 0.617 & 3.69  & 3.505 & 24.8  & 40   & 20 & 8 &  1 Feb 2028 \\ \hline \hline

 Conflicting   & 2016 CL18      & 1.366 & 0.434 & 0.41  & 4.733 & 24.0  & 50   & 9  & 9 &   20 Dec 2026 \\  \hline \hline

               & 2001 FB90      & 2.455 & 0.782 & 1.92  & 2.976 & 19.8  & 380  & 19 & 9 &  8 Sep 2028 \\ \cline{2-11}
               & 2001 VE76      & 1.744 & 0.515 & 4.17  & 3.974 & 23.6  & 70   & 16 & 9 & 5 Feb 2027 \\ \cline{2-11}
               & 2004 JP12      & 1.908 & 0.778 & 8.26  & 3.480 & 23.4  & 70   & 32 & 9 &  28 Nov 2024 \\ \cline{2-11}
               & 2012 TQ231     & 1.195 & 0.465 & 3.36  & 5.200 & 27.6  & 10   & 20 & 9 &  12 May 2024 \\ \cline{2-11}
 Low           & 2015 FA345     & 2.363 & 0.581 & 4.70  & 3.295 & 24.1  & 50   & 15 & 9 &  10 Sep 2025 \\ \cline{2-11}
               & 2017 FA159     & 1.548 & 0.400 & 0.53  & 4.362 & 28.5  & 7    &  8 & 8 &  7 Jan 2024 \\ \cline{2-11}
               & 2020 BO9       & 2.116 & 0.619 & 5.44  & 3.456 & 23.5  & 70   & 19 & 8 &  9 May 2026 \\ \cline{2-11}
               & 2020 HX2       & 0.860 & 0.313 & 9.55  & 6.810 & 21.9  & 140  & 8  & 7 &  9 Oct 2026 \\ \cline{2-11}
               & 2020 US7       & 1.078 & 0.189 & 7.28  & 5.714 & 25.8  & 20   & 6  & 9 & 9 Nov 2024\\ \hline \hline

               & 373135         & 1.652 & 0.497 & 4.39  & 4.124 & 19.5  & 1050 & 9  & 0 & 15 Dec 2023 \\ \cline{2-11}
 Other         & 2014 AD16      & 1.401 & 0.368 & 0.35  & 4.677 & 27.4  & 10   & 14 & 4 & 13 Jul 2027\\ \cline{2-11} 
               & 25143 Itokawa  & 1.324 & 0.280 & 1.62  & 4.898 & 19.3  & 330  & 7  & 0 & 12 Oct 2025 \\ \hline

    \end{tabular}
    \caption{Summary of encounter circumstances of known asteroids with 99942 Apophis during the time frame of this study. Orbital and physical data is from the JPL Small-Body Database. The relative velocity $v_{rel}$ is measured at the MOID. ``Encounter date'' is the date that Apophis is at its MOID with the nominal asteroid orbit. The actual time of any encounter with material on the asteroid orbit will depend on the orbital uncertainty, see text for more details. For diameter estimates, the albedo is assumed to be 0.15 unless it is otherwise available. \label{tab:candidates}}
\end{table}

\section{Conclusions}

We have detailed the encounter circumstances of asteroid 99942 Apophis
with other asteroids and comets in the catalogues of known Solar
System objects, taking full account of the observational uncertainties
on the orbit determinations. No cases likely to result in direct
collisions of known small bodies with Apophis were found. Instances
where there is some risk of impact on Apophis by material accompanying
known small bodies were identified. These objects could benefit from
additional study to confirm or deny the presence of material along
their orbits. Times of close approach of asteroids with Apophis are
listed and the observing circumstances detailed to assist with
observational monitoring of those times at which an impact with debris
might occur, to maintain situational awareness.

We also identified a number of known asteroids whose encounter
circumstances could be better assessed if more recovery or precovery
observations were available, as the quality of an orbit determination
is directly related to the time span of observations available. We
stress however that the shortage of observations of these objects does
not indicate any lack of effort on the part of the global astronomical
community to track these objects: astronomers both professional and
amateur alike often go to great lengths to observe minor bodies under
very difficult conditions. But small bodies can often only be
telescopically observed for a brief period after discovery, as they
move away from us and become too faint, move into the daytime sky
where the glare of the Sun prevents them being observed, or as a
result of poor weather at observing sites. 

The likelihood of a known small Solar System body (or any material
released therefrom) colliding with Apophis is extremely low. The most
likely outcome of careful monitoring of the encounters between Apophis
and the asteroids and comets discussed in this work is that no impact
will be observed. The small probability of collision is however
counter-balanced by disproportionately large consequences. Because of
the hazard associated with even a small perturbation to this
Earth-threatening asteroid, there is ample motivation to determine the
risk as precisely as possible, and we encourage future efforts in that
direction.

\begin{acknowledgements}
We thank the JPL Solar System Dynamics group and the NeoDys Near-Earth
Object Dynamics consortium for providing the orbital determinations used in this
work. We express our appreciation to observers both professional and
amateur who have contributed to the worldwide effort of cataloguing
near-Earth objects. This work was supported in part by the NASA
Meteoroid Environment Office under Cooperative Agreement No.
80NSSC21M0073, and by the Natural Sciences and Engineering Research
Council of Canada (NSERC) Discovery Grant program (grant
No. RGPIN-2018-05659).
\end{acknowledgements}

\bibliography{Wiegert}{}
\bibliographystyle{aasjournal}



\end{document}